\documentclass[prb,twocolumn,showpacs,superscriptaddress,floatfix]{revtex4-1}
\usepackage{graphicx,amsfonts,amssymb,amsmath,xspace,dsfont}
\usepackage[colorlinks=true,citecolor=green,linkcolor=red,urlcolor=blue]{hyperref}
\usepackage{color}
\definecolor{g}{rgb}{.1,0.4,.1} % {.0,0.7,.5}
\definecolor{b}{rgb}{0,0.2,1}
\definecolor{rouge}{rgb}{0.82,0.,0.}
\definecolor{vert}{rgb}{0.,0.82,0.}
\definecolor{orange}{rgb}{1,0.5,0.}
\definecolor{bleu}{rgb}{0.,0.,0.82}
\definecolor{m}{rgb}{0.82,0.,0.82}
\definecolor{vert2}{rgb}{0.,0.5,0.}
\definecolor{rougeclair}{rgb}{1.0,0.7,0.7}

\newcommand{\br}{\boldsymbol r}
\newcommand{\bd}{\boldsymbol d}

\newcommand{\bmm}{\boldsymbol m}

\newcommand{\rmi}{{\rm i}}
\usepackage{overpic}

\begin{document}

\title{Spectroscopy of a topological phase}

\author{Michael Kamfor}
\email{michael.kamfor@tu-dortmund.de}
\affiliation{Lehrstuhl f\"{u}r Theoretische Physik I, Otto-Hahn-Stra\ss e 4, TU Dortmund, 44221 Dortmund, Germany}
\affiliation{Laboratoire de Physique Th\'eorique de la Mati\`ere Condens\'ee, CNRS UMR 7600, \\ 
Universit\'e Pierre et Marie Curie, 4 Place Jussieu, 75252 Paris Cedex 05, France}

\author{S\'ebastien Dusuel}
\email{sdusuel@gmail.com}
\affiliation{Lyc\'ee Saint-Louis, 44 Boulevard Saint-Michel, 75006 Paris, France}

\author{Julien Vidal}
\email{vidal@lptmc.jussieu.fr}
\affiliation{Laboratoire de Physique Th\'eorique de la Mati\`ere Condens\'ee, CNRS UMR 7600, \\ 
Universit\'e Pierre et Marie Curie, 4 Place Jussieu, 75252 Paris Cedex 05, France}

\author{Kai Phillip Schmidt}
\email{kai.schmidt@tu-dortmund.de}
\affiliation{Lehrstuhl f\"{u}r Theoretische Physik I, Otto-Hahn-Stra\ss e 4, TU Dortmund, 44221 Dortmund, Germany}

\begin{abstract}
Dynamical correlation functions of the toric code in a uniform magnetic field are studied inside the topological phase, in the small-field limit. Such an experimentally measurable quantity displays  rich field-dependent features that can be understood via the interplay of the kinetics and the interaction of the anyonic excitations. In particular, it is sensitive to the two-quasiparticle bound states that are present in the spectrum for a wide range of magnetic fields. Interestingly, such collective modes can even constitute the lowest-energy excitations of the system.
\end{abstract}


\maketitle
%%%%%%%%%%%%%%%%%%%%%%%%%%%%%%%%%%%%%%%%%%%%%%
%%%%%%%%%%%%%%%%%%%%%%%%%%%%%%%%%%%%%%%%%%%%%%
%%%%%%%%%%%%%%%%%%%%%%%%%%%%%%%%%%%%%%%%%%%%%%

%
%
%%%%%%%%%%%%%%%%%%%%%%%
%%%%%%%%%%%%%%%%%%%%%%%
\section{Introduction}
%%%%%%%%%%%%%%%%%%%%%%%
%%%%%%%%%%%%%%%%%%%%%%%
%
%
Topologically ordered quantum systems have attracted a tremendous amount of interest in various research fields (see Ref.~\onlinecite{Wen12} for a recent review). In contrast to conventional phases of matter, topologically ordered phases cannot be described by a local order parameter but are rather characterized by non local properties such as a topology-dependent ground-state degeneracy.\cite{Wen89_2} Moreover, elementary excitations in topological phases are particles, called anyons, that obey nontrivial braiding statistics.\cite{Leinaas77,Wilczek82}
Although, theoretically, several models are known to be topologically ordered,  nonambiguous experimental evidence is still lacking. Obviously, changing the topology of a system and measuring its ground-state degeneracy is not an easy task. Thus, a natural issue that arises is to find a measurable smoking-gun signature of topological order. To our knowledge, this remains a challenge but one may already wonder what characteristics of a topologically ordered system are experimentally accessible.

So far, most studies have focused on static low-energy properties of the spectrum. Nevertheless, dynamical properties are known to provide much more informations and are of direct relevance for spectroscopy experiments based, for instance, on inelastic neutron scattering. To bridge the gap between theory and possible future experiments,
dynamical correlation functions have very recently been computed in two different topologically ordered systems.\cite{Punk13,Knolle13}

In this paper, we address similar issues by investigating the toric code\cite{Kitaev03} in a magnetic field. Although the toric code is likely the simplest model displaying topological order, the presence of a uniform field gives rise to unusual transitions in the phase diagram. The latter has been derived by analyzing the ground-state energy as well as the one-quasiparticle (1QP) gap.\cite{Trebst07,Hamma08,Vidal09_1,Vidal09_2,Dusuel11,Tupitsyn10,Wu12}
Computing spin-spin correlation functions is more involved since local spin operators create or annihilate QPs by pairs.  Therefore, one must go beyond 1QP properties. 
Here, we calculate a typical dynamical correlation function inside the topological phase of the perturbed toric code by means of high-order series expansions in the small-field limit. We restrict our computation to 2QP physics which contains most of the spectral weight for sufficiently small magnetic fields.  
This correlation function displays fascinating features originating from the interplay of the anyon kinetics and interactions. In particular, for certain field ranges, strong interactions induce bosonic 2QP bound states, which are the lowest excitations  and dominate the low-energy behavior of the dynamical correlation function.

%
%
%%%%%%%%%%%%%%%%%%%%%%%
%%%%%%%%%%%%%%%%%%%%%%%
\section{Model}
\begin{figure*}[t]
\centering
\includegraphics[width=\textwidth]{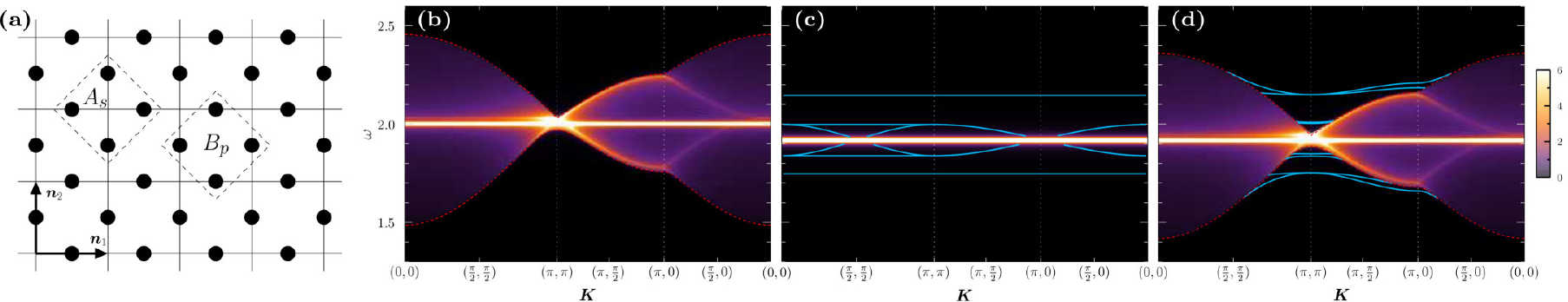}
\caption{\label{fig:DOS}(Color online) The lattice and the unit vectors are shown in (a). Normalized 2QP densities of states (DOSs) as a function of the momentum ${\boldsymbol K}$ for $h_x=0$ and (b) $(h_y,h_z)=(0,0.06)$, (c) $(h_y,h_z)=(0.2,0)$, and (d) $(h_y,h_z)=(0.2,0.06)$. Dashed (red) lines are the boundaries of the continuum computed from the single quasicharge dispersion. Since bound states have a vanishing weight in the DOSs,  they are explicitly highlighted by gray (cyan) lines.
}
\end{figure*}

%%%%%%%%%%%%%%%%%%%%%%%
%%%%%%%%%%%%%%%%%%%%%%%
%
%
Let us consider the Hamiltonian of the toric code\cite{Kitaev03} in a  uniform magnetic field \mbox{$\boldsymbol{h}=(h_x,h_y,h_z)$}
% 
%
%%%%%%%%%%%%%%
\begin{equation}
 \label{eq:ham}
 H = - \frac{1}{2}\sum_{s} A_s  - \frac{1}{2} \sum_{p} B_p- \boldsymbol{h}\cdot\sum_{i} {\boldsymbol \sigma}_i,
\end{equation} 
%%%%%%%%%%%%%%
%
%
where ${\boldsymbol \sigma}_i=(\sigma_i^x,\sigma_i^y,\sigma_i^z)$ are the usual Pauli matrices on site $i$. The so-called charge and flux operators are defined as $A_s=\prod_{i \in s} \sigma_i^x$ and $B_p =\prod_{i \in p} \sigma_i^z$, respectively, where $s$  and $p$ refer to stars  and plaquettes of a square lattice [see Fig.~\ref{fig:DOS}(a)]. In the following, we consider open boundary conditions (infinite plane) and, without loss of generality, we restrict our study to $h_\alpha \geqslant 0$.

For $\boldsymbol{h}=0$, one has $[H,A_s]=[H,B_p]=0$ so that the ground state obeys $A_s|0\rangle_0=B_p|0\rangle_0=|0\rangle_0$. Acting on the ground state with $\sigma_i^z$ ($\sigma_i^x$) yields an eigenstate of energy $+2$ with two static excitations existing on the two stars (plaquettes) sharing site $i$, called charges (fluxes). Charges and fluxes are hard-core bosons with mutual semionic statistics.\cite{Kitaev03}
The operator $\sigma_i^y=\mathrm{i}\sigma_i^x\sigma_i^z$, when acting on $|0\rangle_0$, obviously creates both a pair of charges and a pair of fluxes with a total energy $+4$.
In the presence of a magnetic field, charges and fluxes become dispersive and interact. Excitations can then be described in terms of quasicharges and quasifluxes.\cite{Vidal09_1, Vidal09_2, Dusuel11}

%
%%%%%%%%%%%%%%%%%%%%%%%
%%%%%%%%%%%%%%%%%%%%%%%
%

%
%
%%%%%%%%%%%%%%%%%%%%%%%
%%%%%%%%%%%%%%%%%%%%%%%
\section{Dynamical correlation functions}
%%%%%%%%%%%%%%%%%%%%%%%
%%%%%%%%%%%%%%%%%%%%%%%
%
%
To probe the spectral and dynamical properties of the system, natural quantities are  the space- and time-dependent correlation functions. For simplicity, we focus on the function  $\mbox{}_{\boldsymbol{h}}\langle 0| \sigma_i^z(t) \sigma_j^z(0)|0\rangle_{\boldsymbol{h}}$ whose Fourier transform reads
%
%%%%%%%%%%%%%%%%%%%%%%%
\begin{equation}
\label{eq:S}
\mathcal{S} ({\boldsymbol K},\omega) = \lim_{\epsilon\to 0^+} -\frac{1}{N\pi}\mathrm{Im}\sum_{n} \frac{|A_n({\boldsymbol K})|^2}{\omega-E_n+E_0+\mathrm{i}\,\epsilon},
\end{equation}
%%%%%%%%%%%%%%%%%%%%%%%
%
where \mbox{$A_n({\boldsymbol K})=\sum_{{\boldsymbol r_i}} \mathrm{e}^{\mathrm{i} {\boldsymbol K}\cdot{\boldsymbol r_i}} \mbox{}_{\boldsymbol{h}}\langle n|\sigma_i^z|0\rangle_{\boldsymbol{h}}$}  ($|n\rangle_{\boldsymbol{h}}$ denotes the $n^{\rm th}$ excited state of $H$ with eigenenergy $E_n$)  and $N$ is the number of unit cells.
In addition, we  consider only the case where the observable $\sigma_i^z$ acts on vertical links. As can be checked, the total spectral weight obeys the following sum rule:
%
%%%%%%%%%%%%%%%%%%%%%%%
\begin{equation}
\label{eq:sumrule}
W =\int_{\rm 1BZ} \frac{\mathrm{d}{\boldsymbol K}}{(2\pi)^2}
\int^{+\infty}_{-\infty} \hspace{-3mm}{\mathrm d} \omega \: \mathcal{S} ({\boldsymbol K},\omega)=1,
\end{equation}
%%%%%%%%%%%%%%%%%%%%%%%
%
%
where ${\rm 1BZ}$ stands for the first Brillouin zone.

For $\boldsymbol{h}=0$, as explained above, $\sigma_i^z|0\rangle_{0}$ is an eigenstate of $H$ with two (static) charges so that $\mathcal{S} ({\boldsymbol K},\omega)=\delta(\omega-2)$.
The situation changes when a magnetic field is switched on since in this case QPs interact and become dispersive, so that $\mathcal{S}$ displays more interesting features. However, for small fields, most of the spectral weight remains in the 2QP channel near $\omega=2$.
%
%
%%%%%%%%%%%%%%%%%%%%%%%
\begin{figure*}[t]
\centering
\includegraphics[width=\textwidth]{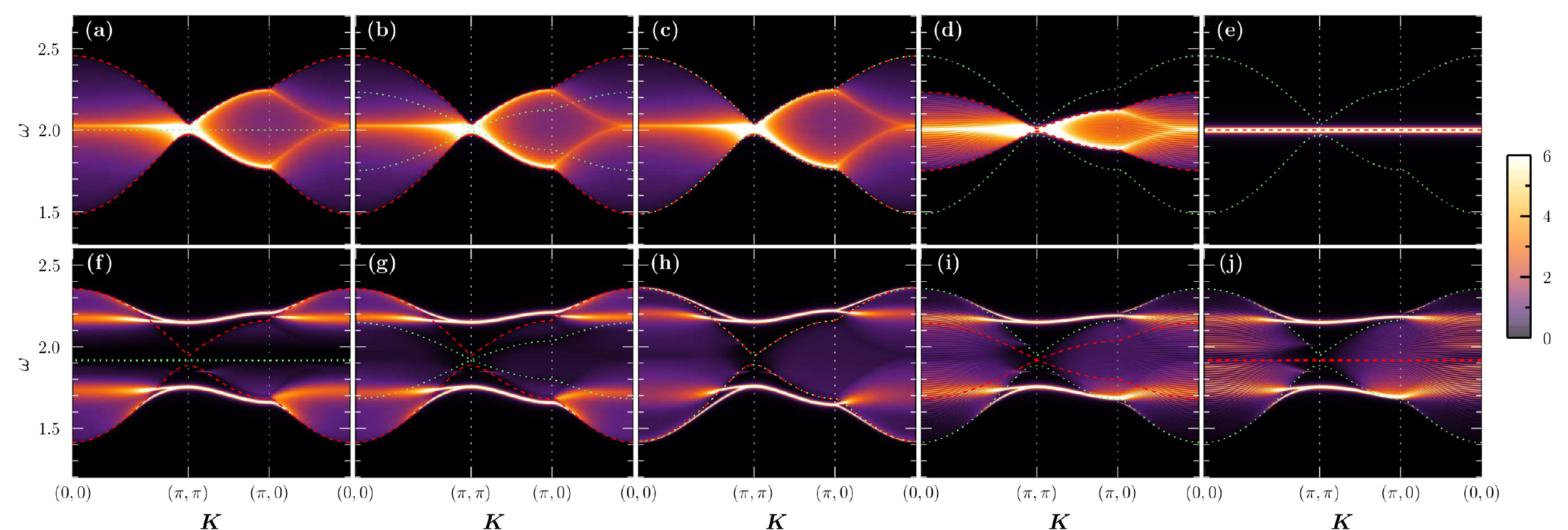}
\caption{\label{fig:S8}(Color online)  $\mathcal{S}({\boldsymbol K},\omega)$ in the 2QP approximation.
Upper and lower panels correspond to $h_y=0$ and \mbox{$h_y=0.2$}, respectively. From left to right $(h_x,h_z)=$ $(0,0.06)$, $(0.03,0.06)$, $(0.06,0.06)$, $(0.06,0.03)$, and $(0.06,0)$. Boundaries of quasicharge and quasiflux continua are represented by dashed (red) and dotted (green) lines, respectively.}
\end{figure*} 
%%%%%%%%%%%%%%%%%%%%%%%
%
%

As a first step, let us thus discuss the 2QP spectrum that appears explicitly in Eq.~(\ref{eq:S}). To compute this spectrum, we use perturbative continuous unitary transformations\cite{Wegner94, Knetter00} which is a well suited method to tackle this problem\cite{Knetter01} as already shown in the study of the 0QP and 1QP spectra.\cite{Vidal09_1,Vidal09_2,Dusuel11} In a few words, the idea is to consider an effective, unitarily transformed, Hamiltonian \mbox{$H_\mathrm{eff} = U(\boldsymbol{h}) H U^\dagger (\boldsymbol{h})$} that is block-diagonal in the canonical eigenbasis of the zero-field Hamiltonian. The perturbative continuous unitary transformations technique allows one to compute $H_\mathrm{eff}$ perturbatively in $\boldsymbol{h}$. However, in each block with $q$-particles, once the matrix elements of $H_\mathrm{eff}$ are obtained in the canonical basis, one still has to diagonalize a $q$-body problem. This procedure yields the eigenstates $|n\rangle_0$ of $H_\mathrm{eff}$, related to eigenstates of $H$ through $|n\rangle_0 = U(\boldsymbol{h}) |n\rangle_{\boldsymbol{h}}$.  In the 2QP sector, this diagonalization is performed numerically for large system sizes  ($\sim 10^4\times10^4$ sites) and for any value of the total momentum $\boldsymbol K$ of the two QPs.  Let us underline that $H_\mathrm{eff}$ conserves the parity of the number of quasicharges and of quasifluxes (independently), so that the 2QP sector splits into two subsectors corresponding to  (i) one quasicharge and one quasiflux; (ii) two quasicharges or two quasifluxes. In the present problem, we  consider only the latter subsector since $\sigma_{i,\mathrm{eff}}^z$ creates a superposition of states with an even number of quasicharges and of quasifluxes when acting on $|0\rangle_{0}$ (see the discussion below). Typical densities of states (DOSs) are displayed in Fig.~\ref{fig:DOS} for $h_x=0$. 

For $\boldsymbol{h}=(0,0,h_z)$, the quasicharges become mobile but no binding effect is expected [see Fig.~\ref{fig:DOS}(b)]. Thus, although quasicharges are hard-core bosons, the 2QP excitation energies are qualitatively given by the sum  \mbox{$\varepsilon(\boldsymbol{K}/2-\boldsymbol{q})+\varepsilon(\boldsymbol{K}/2+\boldsymbol{q})$} 
of the two single quasicharge dispersions and give rise to a continuum at fixed ${\boldsymbol K}$. At order 3 (see Ref.~\onlinecite{Kamfor13} for higher orders), for an arbitrary field, this dispersion reads 
%
%%%%%%%%%%%%%%%%%%%%%%%
\begin{eqnarray}
\label{eq:1qpdisp}
\varepsilon(\boldsymbol{k})&=&1-2 h_z \,\alpha(\boldsymbol{k})  -h_y^2+2 h_z^2 \,\Big[2-\alpha(\boldsymbol{k})^2\Big] \\
&&+h_z\,\alpha(\boldsymbol{k}) \left\{h_x^2+\frac{11}{8} h_y^2+2h_z^2 \Big[5-2\alpha(\boldsymbol{k})^2\Big]\right\}, \nonumber
\end{eqnarray}
%%%%%%%%%%%%%%%%%%%%%%%
%
where $\alpha(\boldsymbol{k})=\cos k_1+\cos k_2$, for a single-particle momentum $\boldsymbol{k}=(k_1,k_2)$.
Details inside the continuum can also be qualitatively understood thanks to the 1QP dispersion. In particular, the maxima of the DOSs at fixed $\boldsymbol{K}$ can be understood from the saddle points in $\boldsymbol{q}$ of these 2QP energies. 
Unlike quasicharges, for $h_x=h_y=0$, quasifluxes remain static, yielding a flat band at $\omega=2$. 

For $\boldsymbol{h}=(0,h_y,0)$, the situation is different. Indeed, although a single quasicharge (or quasiflux) remains static,  $\sigma^y$ induces an unusual dynamics of particle pairs (see Ref.~\onlinecite{Vidal09_2} as well as Refs.~\onlinecite{Xu04, Xu05} for closely related issues). The associated dimensional reduction leads to either flat bands or one-dimensional dispersive bands in the 2QP spectrum. Consequently, for a fixed momentum ${\boldsymbol K}$, no continuum is observed and the spectrum is discrete [see Fig.~\ref{fig:DOS}(c)].  As discussed in Ref.~\onlinecite{Vidal09_2}, the corresponding eigenstates are bound states.

When $h_z$ and $h_y$ are finite, the spectrum displays both a quasicharge continuum and some discrete energy levels associated to the aforementioned bound states [see Fig.~\ref{fig:DOS}(d)]. Interestingly, one needs a finite value of $h_y$  to observe bound states out of the continuum. If $h_y$ is large enough, such bound states can even exist for any value of the momentum ${\boldsymbol K}$ and become the lowest energy 2QP states (see discussion below). 
Finally, note that for $h_x=0$, the quasiflux continuum also exists but its width is of order $h_y^2 h_z^2$ (whereas the width of the quasicharge continuum is of order $h_z$). This quasiflux continuum thus looks like a flat band.

The next step to obtain $\mathcal{S} ({\boldsymbol K},\omega)$ is to compute
%
%%%%%%%%%%%%%%%%%%%%%%%
\begin{equation}
\label{eq:matelem}
\mbox{}_{\boldsymbol{h}}\langle n|\sigma_i^z|0\rangle_{\boldsymbol{h}} = 
\mbox{}_{0}\langle n|U(\boldsymbol{h})\sigma_i^z U^\dagger (\boldsymbol{h})|0\rangle_{0}=
\mbox{}_{0}\langle n|\sigma_{i,\mathrm{eff}}^z|0\rangle_{0}.
\end{equation}
%%%%%%%%%%%%%%%%%%%%%%%
%
 As motivated above, for sufficiently small fields, the most relevant matrix elements are those involving  2QP states. More precisely, the contribution of the 2QP sector to the total spectral weight $W$ can be computed perturbatively. At order 4, it reads
%
%%%%%%%%%%%%%%%%%%%%%%%
\begin{eqnarray}
W_{2\mathrm{QP}} &=& \sum_{n\in 2\mathrm{QP}}\left|\mbox{}_{0}\langle n|\sigma_{i,\mathrm{eff}}^z|0\rangle_{0}\right|^2,\nonumber\\
&=& 1-h_x^2-\frac{3}{8}h_y^2-h_z^2-\frac{13}{2}h_x^4-\frac{3445}{2304}h_y^4-\frac{33}{2}h_z^4 \nonumber \\
&&-\frac{425}{64}h_x^2 h_y^2 -\frac{177}{32}h_y^2 h_z^2 + \frac{13}{8}h_x^2h_z^2.
\end{eqnarray}
%%%%%%%%%%%%%%%%%%%%%%%
%
For all values of the field considered here, most of the spectral weight lies in the 2QP sector since $W_{2\mathrm{QP}}>0.94$. As a consequence, we restrict the sum in 
Eq.~(\ref{eq:S}) to 2QP states and neglect higher-energy sectors.
Typical plots of $\mathcal{S}(\boldsymbol{K},\omega)$ are shown in Fig.~\ref{fig:S8} where, as for the DOSs, we used a finite broadening $\epsilon=0.002$.
In practice, we computed the 2QP spectrum (and thus the energy resolution) at order 6 while matrix elements of $\sigma_{i,\mathrm{eff}}^z$ [and thus $\mathcal{S}(\boldsymbol{K},\omega)$] at order 4 (see the Appendix).

Let us start by discussing the case \mbox{$h_y=0$}. As can be seen in Fig.~\ref{fig:S8} (upper panel), $\mathcal{S}(\boldsymbol{K},\omega)$ is almost independent of $h_x$ but strongly varies with $h_z$. To understand this feature, it is worth considering the lowest non-trivial orders. Indeed, (i) at order 0, the effective observable $\sigma_{i,\mathrm{eff}}^z=\sigma_{i}^z$ creates two quasicharges when acting on the ground state; 
(ii) at order 1, the effective Hamiltonian does not couple quasicharges and quasifluxes that behave as hard-core bosons with nearest-neighbor hoppings.
Higher orders lead to several minor modifications of this picture. On the one hand, in the two quasicharges sector, $|A_n({\boldsymbol K})|^2$ is modified from order 2 only. On the other hand, the two quasifluxes sector remains almost irrelevant. Indeed, although $H_\mathrm{eff}$ couples quasicharges and quasifluxes from order 2, $\sigma_{i,\mathrm{eff}}^z$ starts to create quasifluxes at order 3 only, hence yielding \mbox{order-6} corrections to $|A_n({\boldsymbol K})|^2$.
The main features of $\mathcal{S}(\boldsymbol{K},\omega)$ are obviously related to those of the DOS [see Eq.~(\ref{eq:S})]. In particular, at fixed $\boldsymbol{K}$, the extrema of both quantities are observed at the same energies [see for instance Figs.~\ref{fig:DOS}(b) and \ref{fig:S8}(a)]. However, the flat band at $\omega=2$ due to quasifluxes observed in the DOS is absent in $\mathcal{S}(\boldsymbol{K},\omega)$. Let us also stress that the flat band at $\omega=2$ observed in Fig.~\ref{fig:S8}(e) does not originate from quasifluxes but from quasicharges that are static for $h_z=0$.
 
%
%
%%%%%%%%%%%%%%%%%%%%%%%
\begin{figure*}[t]
\centering
\includegraphics[width=\textwidth]{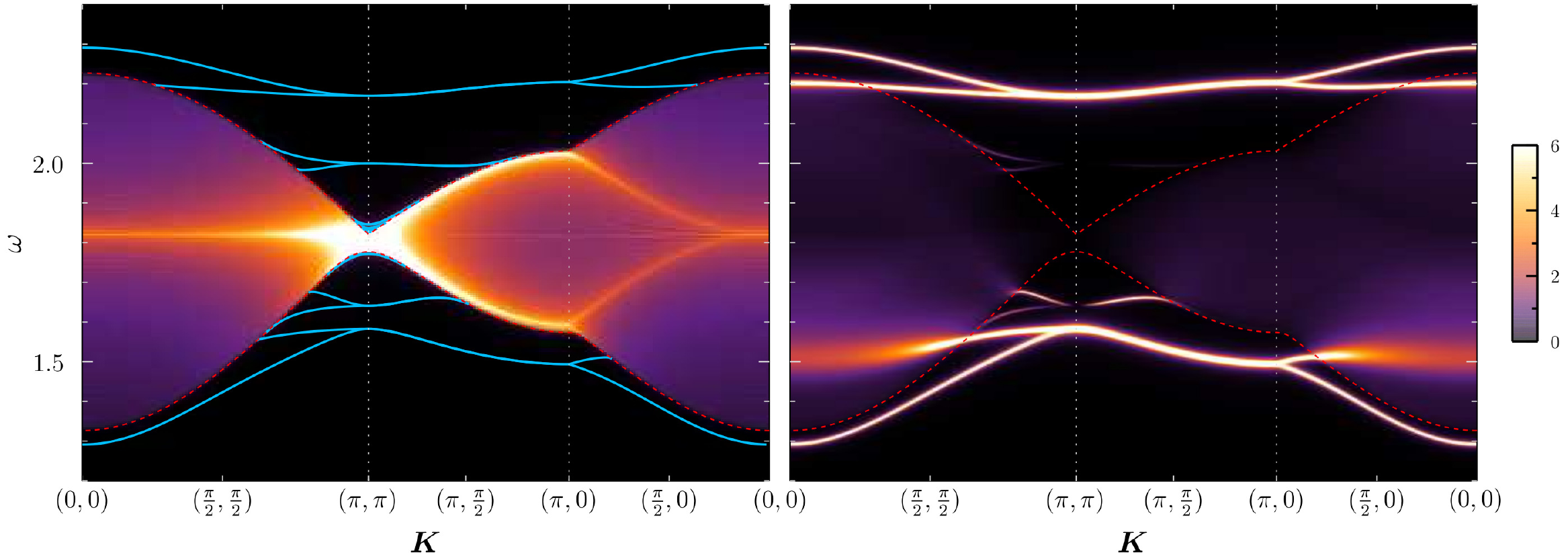}
\caption{\label{fig:comparison}(Color online) Normalized DOS (left) and $\mathcal{S}({\boldsymbol K},\omega)$ (right) for $(h_x,h_y,h_z)=(0.06,0.3,0.06)$ (see Fig.~\ref{fig:DOS} for conventions).}
\end{figure*}
%%%%%%%%%%%%%%%%%%%%%%%
%
%

Drastic changes in $\mathcal{S} ({\boldsymbol K},\omega)$ arise when $h_y\neq 0$ for two reasons: (i) at order 1, quasicharges and quasifluxes interact, giving rise to the previously discussed bound states; (ii) the effective observable generates quasifluxes at the same order.
For sufficiently large $h_y$, an important part of the spectral weight is carried by some of these bound states [see  Fig.~\ref{fig:S8} (lower panel)]. In contrast to the DOS, $\mathcal{S} ({\boldsymbol K},\omega)$ clearly detects these states  and keeps track of them when they merge into the continuum, leaving prominent resonances.
Note also  that the DOS is symmetric under the exchange $h_x \leftrightarrow h_z$ (for any $h_y$) but  $\mathcal{S} ({\boldsymbol K},\omega)$ is not due to the choice of the observable $\sigma^z$.

Let us next comment on the continua boundaries observed in $\mathcal{S} ({\boldsymbol K},\omega)$. For $h_y\neq 0$, these are clearly defined either by the quasicharge dispersion when $h_x<h_z$ [see the dashed (red) lines in Fig.~\ref{fig:S8}(g)] or by the quasiflux dispersion in the opposite case [see the dotted (green) lines in Fig.~\ref{fig:S8}(i)].
The case $h_y=0$ needs to be treated more carefully. In the extreme case $h_z=0$ [see Fig.~\ref{fig:S8}(e)], the previously discussed flat band at $\omega=2$ contains all the spectral weight. Indeed, with $\sigma^x$ as the only perturbation, it is not possible to transmute (virtually) two charges into two fluxes. When $h_z < h_x$ [see Fig.~\ref{fig:S8}(d)], the boundaries of the continuum are actually given by the quasiflux dispersion even if this is not visible with the intensity scale used in the plots. The leading contribution to $|A_n({\boldsymbol K})|^2$ between the dashed (red) and dotted (green) lines is of order $h_x^2 h_z^2$, whereas for $h_y\neq 0$ it is of order $h_y^2$.

Finally, if $h_y$ becomes much larger than $h_x$ and $h_z$ (while remaining in the topological phase), the 2QP extremal-energy states are bound states for any value of the total momentum, as illustrated in Fig.~\ref{fig:comparison}. The lowest-energy excitation is therefore a two-particle bound state at ${\boldsymbol K}=(0,0)$.
A simple way to understand this phenomenon is to consider the limit $h_x=h_z=0$. Indeed, in this case, the spectrum displayed in Fig.~\ref{fig:DOS}(c)  is made only of bound states.\cite{Vidal09_2} The lowest 2QP-energy level is two-fold degenerate whereas, at any finite order, the central band is infinitely degenerate. Switching on $h_x$ and $h_z$ induces two different effects. First, the central band degeneracy is lifted and gives rise to the continuum. Second, the lowest flat band splits  
into two dispersive bands as clearly observed in Fig.~\ref{fig:comparison}. When $h_x$ and $h_z$ are further increased, the bound state eventually merges into the continuum which broadens.  

%
%
%%%%%%%%%%%%%%%%%%%%%%%
%%%%%%%%%%%%%%%%%%%%%%%
\section{Conclusion}
%%%%%%%%%%%%%%%%%%%%%%%
%%%%%%%%%%%%%%%%%%%%%%%
%
%
To date, most studies of topological phases have focused on low-energy properties (the ground state and first excited states). In the present work, we went one step beyond by computing a typical zero-temperature dynamical correlation function in one of the simplest topologically ordered systems, namely, the toric code, in the presence of a magnetic field. This challenging problem requires dealing with high-energy states, but is crucial for potential future experiments.
Using a perturbative approach and a controlled (2QP) approximation, we have unveiled the importance of bound states of anyons that, depending on the magnetic field, can carry most of the spectral weight. The role of these bound states in the breakdown of the topological phase is an interesting question that is still to be elucidated.

Recently, dynamical correlation functions have been computed\cite{Knolle13} in Kitaev's honeycomb model  which is deeply related to the toric code in some limiting cases.\cite{Kitaev06}
A very interesting question would be to analyze the influence of a magnetic field in this system to determine whether bound states also arise in the high-energy spectrum, especially since ultracold atom experiments  may have a chance to simulate this model.\cite{Jiang08}
Finally, studying the effect of temperature on these correlation functions is undoubtedly the next step in the understanding of topological phases of matter, and we hope that the present work will stimulate further investigations in this direction.

%%%%%%%%%%%%%%%%%%%%%%%%%
%%%%%%%%%%%%%%%%%%%%%%%%%
%                       Acknowledgments                      %
%%%%%%%%%%%%%%%%%%%%%%%%%
%%%%%%%%%%%%%%%%%%%%%%%%%
\acknowledgements
K.P.S. and M.K. acknowledge ESF and EuroHorcs for funding through the EURYI as well as DFG.

%%%%%%%%%%%%%%%%%%%%%%%%%
%%%%%%%%%%%%%%%%%%%%%%%%%
%                       Appendices                    %
%%%%%%%%%%%%%%%%%%%%%%%%%
%%%%%%%%%%%%%%%%%%%%%%%%%
\appendix

%
%%%%%%%%%%%%%%%%%%%%%%%
%%%%%%%%%%%%%%%%%%%%%%%
\section{2QP matrix elements of the effective Hamiltonian $H_{\rm eff}$}
%%%%%%%%%%%%%%%%%%%%%%%
%%%%%%%%%%%%%%%%%%%%%%%
%
\label{App:A}

\begin{figure}[h]
	\begin{minipage}[c]{0.96\columnwidth}
	\includegraphics[width=0.96\columnwidth]{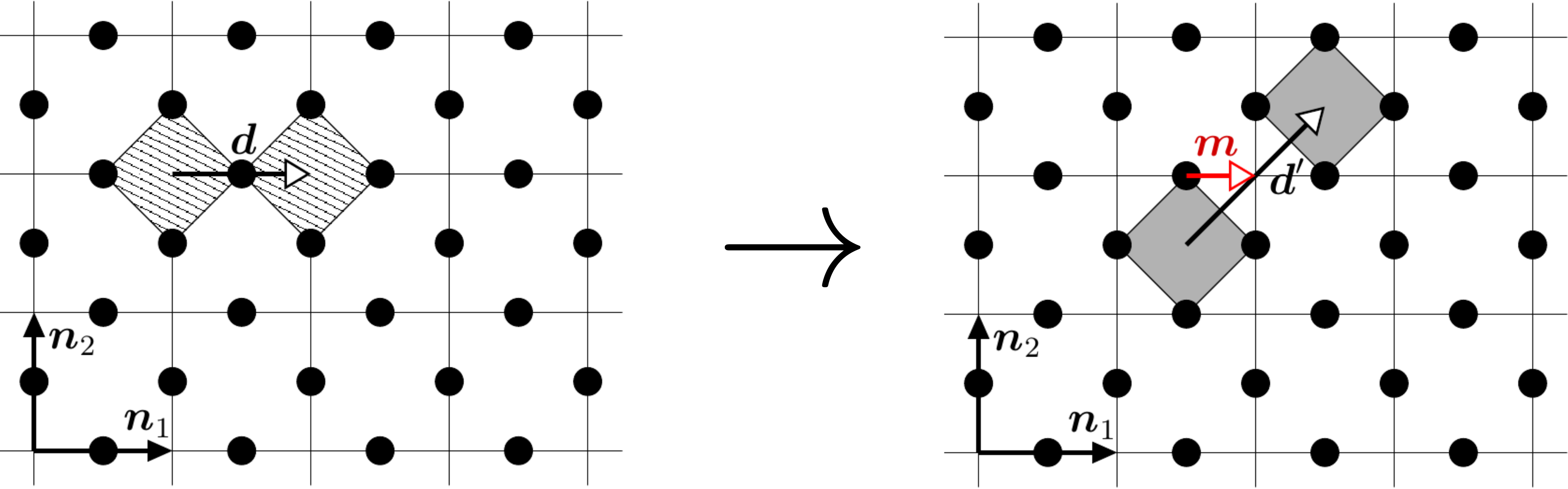}
	\end{minipage}	
\caption{(Color online) A typical hopping process with an initial state (left) that consists in two quasicharges ($\tau={\rm c}$), separated by a vector $\bd=(1,0)$ and a final state (right) that consists in two quasifluxes ($\tau'={\rm f}$), separated by a vector $\bd'=(1,1)$. This hopping induces a shift  ${\boldsymbol m}=(\frac{1}{2},0)$ of the center of mass. The corresponding hopping amplitude is $- \frac{\rmi}{2} h_x h_y$ (see Table \ref{tab:heffcoeff}).}
\label{fig:heff2}
\end{figure}

In this appendix, we give 2QP matrix elements of $H_{\rm eff}$ in the two-quasicharge or two-quasiflux subsector, up to order 2 (series up to order 6 are available upon request). These elements are denoted by $t_{\tau'\!\!,\bd'\!;\,{\boldsymbol m}}^{\tau\!\,,\bd\:\!}$. Initial and final pair types (${\rm c}$ for charges and ${\rm f}$ for fluxes) are given by $\tau$ and $\tau'$, respectively. The hopping vector of the center of mass is denoted by ${\boldsymbol m}$. The vector $\bd$ ($\bd'$) gives the relative positions of the initial (final) particles. Since the two particles are indistinguishable (hard-core bosons), one must find a proper way to encode the corresponding states. Here and hereafter, one chooses the coordinates of the vector $\bd=(d_1,d_2)$ in the basis $(\boldsymbol{n}_1,\boldsymbol{n}_2)$, such that either $d_1>0$ or, if $d_1=0$, $d_2 > 0$. These notations are illustrated in Fig.~\ref{fig:heff2}.
Table \ref{tab:heffcoeff} contains matrix elements that are not related by lattice symmetries and/or Hermiticity. Furthermore we consider only initial states that are quasicharges since one can recover the matrix elements for two initial quasifluxes by  exchanging $h_x$ and $h_z$ (charge-flux symmetry).

The matrix elements discussed above capture the interactions between the QPs. These must be combined with 1QP hopping terms to obtain the effective Hamiltonian of two QPs. Diagonalization of this Hamiltonian gives 2QP excitation energies. To be concrete, let us consider the hopping of a quasicharge with vector $p_1\boldsymbol{n}_1 + p_2\boldsymbol{n}_2$, whose amplitude is denoted $t_{p_1,p_2}$. Up to order 2 (series up to order 8 are available upon request), one has
$t_{0,0} = 1-h_y^2+2h_z^2$, \, $t_{1,0} = -h_z$, \, $t_{1,1} = -h_z^2$, and $t_{2,0}=-\frac{1}{2}h_z^2$. As before, one should use lattice symmetries and Hermiticity to recover all possible hopping amplitudes of quasicharges, and charge-flux symmetry to get those of quasifluxes. These amplitudes allows one to compute the order-2 dispersions of quasicharges and quasifluxes.
%
%
%\begin{figure}[h]
%	\begin{minipage}[c]{0.44\columnwidth}
%	\includegraphics[width=0.75\columnwidth]{./fig4}
%	\end{minipage}	
%	\hfill
%	\begin{minipage}[c]{0.08\columnwidth}
%	{\huge $\rightarrow$}
%	\end{minipage}
%	\hfill
%	\begin{minipage}[c]{0.44\columnwidth} 
%	\includegraphics[width=0.75\columnwidth]{./fig5}
%	\end{minipage}
%\caption{A typical hopping process with an initial state (left) that consists in two quasicharges ($\tau={\rm c}$), separated by a vector $\bd=(1,0)$ and a final state (right) that consists in two quasifluxes ($\tau'={\rm f}$), separated by a vector $\bd'=(1,1)$. This hopping induces a shift  ${\boldsymbol m}=(\frac{1}{2},0)$ of the center of mass. The corresponding hopping amplitude is $- \frac{\rmi}{2} h_x h_y$ (see Table \ref{tab:heffcoeff}).}
%\label{fig:heff2}
%\end{figure}
%
%
\begin{table}[h]
\begin{tabular}{||l|l|l|l|l|l||}\hline
$\tau$ & $\:\:\:\: \boldsymbol d$ & $\tau'$ & $\:\:\:\: \boldsymbol d'$ & $\:\:\:\:\:{\boldsymbol m}$  & $\:\:\:\:\:\:t_{\tau'\!\!,\bd'\!;\,{\boldsymbol m}}^{\tau\!\,,\bd\:\!}\phantom{\Big(}$\\\hline
c &$(1,0) $&f &$(0,1) $&$(0, 0) $&$ - \rmi\,  h_y + h_x h_z$\\
c &$(1, 0) $&f &$(0,2) $&$(0,\frac{1}{2})$ &$-  \frac{\rmi}{2} h_x h_y$\\
c &$(1, 0) $&c &$(1, 0) $&$(0, 0)$ &$4 h_z^2-\frac{5}{4} h_y^2$\\
c &$(1, 0) $&c &$(1, 0) $&$(1, 0)$ &$\frac{1}{2} h_z^2$\\
c &$(1, 0) $&f &$(1, 1)$ &$(\frac{1}{2}, 0)$ &$- \frac{\rmi}{2} h_x h_y$\\
c &$(1, 1) $&f &$(1, 1) $&$(\frac{1}{2}, -\frac{1}{2})$ &$ h_y^2$\\[2pt]\hline
\end{tabular}
\caption{Order-2 2QP matrix elements of $H_{\rm eff}$ that are not related by lattice symmetries and/or Hermiticity [see text and Fig.~\ref{fig:heff2} for notations].}
\label{tab:heffcoeff}
\end{table}

%
%%%%%%%%%%%%%%%%%%%%%%%
%%%%%%%%%%%%%%%%%%%%%%%
\section{Matrix elements of the effective observable $\sigma^z_{{\boldsymbol r},{\rm eff}}$}
\begin{figure}[h]
	\begin{minipage}[b]{0.96\columnwidth}
	\includegraphics[width=0.96\columnwidth]{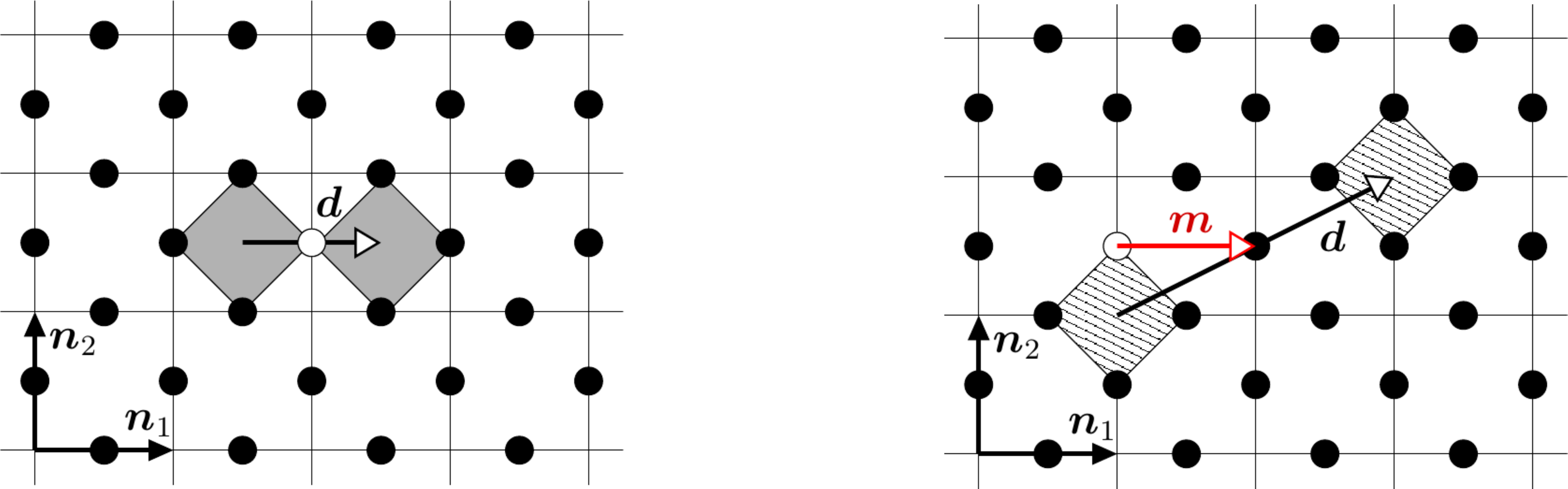}
	\end{minipage}
\caption{Two (among many) states generated by the action of the effective observable $\sigma^z_{ \boldsymbol r,\rm eff}$ on the white site.  Left: $\tau=\,$f, $\bd=(1,0)$, $\boldsymbol{m}=(0,0)$, and the amplitude is $-  \frac{\rmi}{4} h_y$. Right: $\tau=\,$c, $\bd=(2,1)$, $\boldsymbol{m}=(1,0)$, and the amplitude is $\frac{5}{8} h_z^2$.}
\label{fig:obs}
\end{figure}

%%%%%%%%%%%%%%%%%%%%%%%
%%%%%%%%%%%%%%%%%%%%%%%
%
\label{App:B}
The action of the effective observable at position ${\boldsymbol r}$ on the zero-field ground state can be written as
\begin{eqnarray}
 \sigma^z_{ \boldsymbol r,\rm eff} \left|0\right\rangle_0 &=& \sum\limits_{ \tau,\bd,\bmm} A_{\tau,\bd,\bmm} 
 \left| \tau, \br +\bmm-\bd/2,\br +\bmm+\bd/2 \right\rangle \nonumber \\
 && + \cdots
 \label{eq:eff_obs_gs}
\end{eqnarray}
The sum runs over all 2QP states (with two quasifluxes or two quasicharges) whereas ellipsis stand for states that have more than 2QPs. The two indistinguishable particles of type $\tau$ have coordinates $\br+\bmm+\bd/2$ and $\br+\bmm-\bd/2$, where $\bd$ and $\tau$ are defined as in the previous section, and where $\bmm$ is the vector linking position $\br$ to the center of mass of the two particles  (see Fig.~\ref{fig:obs} for illustration). The amplitudes $A_{\tau,\bd,\bmm}$ are given in Table \ref{tab:obscoeff} up to order 2 (series up to order 4 are available upon request), for an action of the effective observable on a vertical link ${\boldsymbol r}$. Amplitudes for horizontal links are straightforwardly obtained from lattice symmetries. Let us finally note that amplitudes for the observable $\sigma^x_{{\boldsymbol r},{\rm eff}}$ can be deduced from the charge-flux symmetry.

%\begin{figure}[h]
%	\begin{minipage}[b]{0.44\columnwidth}
%	\includegraphics[width=0.75\columnwidth]{./fig6}
%	\end{minipage}
%	\hfill
%	\begin{minipage}[b]{0.44\columnwidth}
%	\includegraphics[width=0.75\columnwidth]{./fig7}
%	\end{minipage}
%\caption{Two (among many) states generated by the action of the effective observable $\sigma^z_{ \boldsymbol r,\rm eff}$ on the white site.  Left: $\tau=\,$f, $\bd=(1,0)$, $\boldsymbol{m}=(0,0)$, and the amplitude is $-  \frac{\rmi}{4} h_y$. Right: $\tau=\,$c, $\bd=(2,1)$, $\boldsymbol{m}=(1,0)$, and the amplitude is $\frac{5}{8} h_z^2$.}
%\label{fig:obs}
%\end{figure}

\begin{table}[h]
\begin{tabular}{||l|l|l|l||}\hline
$\tau$ & $\:\:\:\:\boldsymbol d$ &  $\:\:\:\: \boldsymbol m$ &  $\:\:\:\:\:\:\:\:\:\:\:\:\:\:\:A_{\tau, \bd, \bmm}\phantom{\Big(}$ \\\hline
c & $(0, 1)$ & $(0, 0)$ & $1-\frac{1}{2} h_x^2-\frac{7}{32} h_y^2-\frac{5}{4} h_z^2$ \\
c & $(0,2)$ & $(0, \frac{1}{2})$ & $\frac{1}{2} h_z-  \frac{\rmi}{4} h_x h_y$ \\
c & $(1,1)$ & $(\frac{1}{2}, 0)$ & $\frac{1}{2} h_z-  \frac{\rmi}{4} h_x h_y$ \\
f & $(1, 0)$ & $(0, 0)$ & $-\frac{\rmi}{4} h_y$ \\
c & $(0, 1)$ & $(0, 1)$ & $\frac{1}{8} h_z^2$ \\
c & $(0, 1)$ & $(1, 0)$ & $\frac{1}{4} h_z^2$ \\
c & $(0, 3)$ & $(0, 0)$ & $\frac{1}{4} h_z^2$ \\
c & $(0, 3)$ & $(0, 1)$ & $\frac{5}{8} h_z^2$ \\
c & $(1, 0)$ & $(\frac{1}{2}, \frac{1}{2})$ & $\frac{3}{4} h_z^2$ \\
c & $(1, 2)$ & $(\frac{1}{2}, -\frac{1}{2})$ & $\frac{1}{4} h_z^2$ \\
c & $(1, 2)$ & $(\frac{1}{2}, \frac{1}{2})$ & $\frac{5}{4} h_z^2$ \\
c & $(2, 1)$ & $(0, 0)$ & $\frac{1}{4} h_z^2$ \\
c & $(2, 1)$ & $(1, 0)$ & $\frac{5}{8} h_z^2$ \\
f & $(1, 1)$ & $(0, \frac{1}{2})$ & $-  \frac{\rmi}{16} h_x h_y$ \\
f & $(2, 0)$ & $(\frac{1}{2}, 0)$ & $-  \frac{\rmi}{16} h_x h_y$ \\[2pt]\hline
\end{tabular}
\caption{Order-2 amplitudes associated to $\sigma^z_{ \boldsymbol r,\rm eff}$  [see Eq.~(\ref{eq:eff_obs_gs}) and Fig.~\ref{fig:obs} for notations].}
\label{tab:obscoeff}
\end{table}

\newpage

%\bibliography{./biblio_TCSD.bib}

\begin{thebibliography}{22}%
\makeatletter
\providecommand \@ifxundefined [1]{%
 \@ifx{#1\undefined}
}%
\providecommand \@ifnum [1]{%
 \ifnum #1\expandafter \@firstoftwo
 \else \expandafter \@secondoftwo
 \fi
}%
\providecommand \@ifx [1]{%
 \ifx #1\expandafter \@firstoftwo
 \else \expandafter \@secondoftwo
 \fi
}%
\providecommand \natexlab [1]{#1}%
\providecommand \enquote  [1]{``#1''}%
\providecommand \bibnamefont  [1]{#1}%
\providecommand \bibfnamefont [1]{#1}%
\providecommand \citenamefont [1]{#1}%
\providecommand \href@noop [0]{\@secondoftwo}%
\providecommand \href [0]{\begingroup \@sanitize@url \@href}%
\providecommand \@href[1]{\@@startlink{#1}\@@href}%
\providecommand \@@href[1]{\endgroup#1\@@endlink}%
\providecommand \@sanitize@url [0]{\catcode `\\12\catcode `\$12\catcode
  `\&12\catcode `\#12\catcode `\^12\catcode `\_12\catcode `\%12\relax}%
\providecommand \@@startlink[1]{}%
\providecommand \@@endlink[0]{}%
\providecommand \url  [0]{\begingroup\@sanitize@url \@url }%
\providecommand \@url [1]{\endgroup\@href {#1}{\urlprefix }}%
\providecommand \urlprefix  [0]{URL }%
\providecommand \Eprint [0]{\href }%
\providecommand \doibase [0]{http://dx.doi.org/}%
\providecommand \selectlanguage [0]{\@gobble}%
\providecommand \bibinfo  [0]{\@secondoftwo}%
\providecommand \bibfield  [0]{\@secondoftwo}%
\providecommand \translation [1]{[#1]}%
\providecommand \BibitemOpen [0]{}%
\providecommand \bibitemStop [0]{}%
\providecommand \bibitemNoStop [0]{.\EOS\space}%
\providecommand \EOS [0]{\spacefactor3000\relax}%
\providecommand \BibitemShut  [1]{\csname bibitem#1\endcsname}%
\let\auto@bib@innerbib\@empty
%</preamble>
\bibitem [{\citenamefont {Wen}()}]{Wen12}%
  \BibitemOpen
  \bibfield  {author} {\bibinfo {author} {\bibfnamefont {X.-G.}\ \bibnamefont
  {Wen}},\ }\href@noop {} {}\bibinfo {note}
  {\href{http://arxiv.org/abs/1210.1281}{ arXiv:1210.1281}}\BibitemShut
  {NoStop}%
\bibitem [{\citenamefont {Wen}(1989)}]{Wen89_2}%
  \BibitemOpen
  \bibfield  {author} {\bibinfo {author} {\bibfnamefont {X.-G.}\ \bibnamefont
  {Wen}},\ }\href {\doibase 10.1103/PhysRevB.40.7387} {\bibfield  {journal}
  {\bibinfo  {journal} {Phys. Rev. B}\ }\textbf {\bibinfo {volume} {40}},\
  \bibinfo {pages} {7387} (\bibinfo {year} {1989})}\BibitemShut {NoStop}%
\bibitem [{\citenamefont {Leinaas}\ and\ \citenamefont
  {Myrheim}(1977)}]{Leinaas77}%
  \BibitemOpen
  \bibfield  {author} {\bibinfo {author} {\bibfnamefont {J.~M.}\ \bibnamefont
  {Leinaas}}\ and\ \bibinfo {author} {\bibfnamefont {J.}~\bibnamefont
  {Myrheim}},\ }\href {\doibase 10.1007/BF02727953} {\bibfield  {journal}
  {\bibinfo  {journal} {Il Nuovo Cimento B}\ }\textbf {\bibinfo {volume}
  {37}},\ \bibinfo {pages} {1} (\bibinfo {year} {1977})}\BibitemShut {NoStop}%
\bibitem [{\citenamefont {Wilczek}(1982)}]{Wilczek82}%
  \BibitemOpen
  \bibfield  {author} {\bibinfo {author} {\bibfnamefont {F.}~\bibnamefont
  {Wilczek}},\ }\href {\doibase 10.1103/PhysRevLett.49.957} {\bibfield
  {journal} {\bibinfo  {journal} {Phys. Rev. Lett.}\ }\textbf {\bibinfo
  {volume} {49}},\ \bibinfo {pages} {957} (\bibinfo {year} {1982})}\BibitemShut
  {NoStop}%
\bibitem [{\citenamefont {Punk}\ \emph {et~al.}()\citenamefont {Punk},
  \citenamefont {Chowdhury},\ and\ \citenamefont {Sachdev}}]{Punk13}%
  \BibitemOpen
  \bibfield  {author} {\bibinfo {author} {\bibfnamefont {M.}~\bibnamefont
  {Punk}}, \bibinfo {author} {\bibfnamefont {D.}~\bibnamefont {Chowdhury}}, \
  and\ \bibinfo {author} {\bibfnamefont {S.}~\bibnamefont {Sachdev}},\
  }\href@noop {} {}\bibinfo {note}
  {\href{http://arxiv.org/abs/1308.2222}{arXiv:1308.2222}}\BibitemShut
  {NoStop}%
\bibitem [{\citenamefont {Knolle}\ \emph {et~al.}()\citenamefont {Knolle},
  \citenamefont {Kovrizhin}, \citenamefont {Chalker},\ and\ \citenamefont
  {Moessner}}]{Knolle13}%
  \BibitemOpen
  \bibfield  {author} {\bibinfo {author} {\bibfnamefont {J.}~\bibnamefont
  {Knolle}}, \bibinfo {author} {\bibfnamefont {D.~L.}\ \bibnamefont
  {Kovrizhin}}, \bibinfo {author} {\bibfnamefont {J.~T.}\ \bibnamefont
  {Chalker}}, \ and\ \bibinfo {author} {\bibfnamefont {R.}~\bibnamefont
  {Moessner}},\ }\href@noop {} {}\bibinfo {note}
  {\href{http://arxiv.org/abs/1308.4336}{arXiv:1308.4336}}\BibitemShut
  {NoStop}%
\bibitem [{\citenamefont {Kitaev}(2003)}]{Kitaev03}%
  \BibitemOpen
  \bibfield  {author} {\bibinfo {author} {\bibfnamefont {A.~Y.}\ \bibnamefont
  {Kitaev}},\ }\href {\doibase 10.1016/S0003-4916(02)00018-0} {\bibfield
  {journal} {\bibinfo  {journal} {Ann. Phys. (N.Y.)}\ }\textbf {\bibinfo
  {volume} {303}},\ \bibinfo {pages} {2} (\bibinfo {year} {2003})}\BibitemShut
  {NoStop}%
\bibitem [{\citenamefont {Trebst}\ \emph {et~al.}(2007)\citenamefont {Trebst},
  \citenamefont {Werner}, \citenamefont {Troyer}, \citenamefont {Shtengel},\
  and\ \citenamefont {Nayak}}]{Trebst07}%
  \BibitemOpen
  \bibfield  {author} {\bibinfo {author} {\bibfnamefont {S.}~\bibnamefont
  {Trebst}}, \bibinfo {author} {\bibfnamefont {P.}~\bibnamefont {Werner}},
  \bibinfo {author} {\bibfnamefont {M.}~\bibnamefont {Troyer}}, \bibinfo
  {author} {\bibfnamefont {K.}~\bibnamefont {Shtengel}}, \ and\ \bibinfo
  {author} {\bibfnamefont {C.}~\bibnamefont {Nayak}},\ }\href {\doibase
  10.1103/PhysRevLett.98.070602} {\bibfield  {journal} {\bibinfo  {journal}
  {Phys. Rev. Lett.}\ }\textbf {\bibinfo {volume} {98}},\ \bibinfo {pages}
  {070602} (\bibinfo {year} {2007})}\BibitemShut {NoStop}%
\bibitem [{\citenamefont {Hamma}\ and\ \citenamefont {Lidar}(2008)}]{Hamma08}%
  \BibitemOpen
  \bibfield  {author} {\bibinfo {author} {\bibfnamefont {A.}~\bibnamefont
  {Hamma}}\ and\ \bibinfo {author} {\bibfnamefont {D.~A.}\ \bibnamefont
  {Lidar}},\ }\href {\doibase 10.1103/PhysRevLett.100.030502} {\bibfield
  {journal} {\bibinfo  {journal} {Phys. Rev. Lett.}\ }\textbf {\bibinfo
  {volume} {100}},\ \bibinfo {pages} {030502} (\bibinfo {year}
  {2008})}\BibitemShut {NoStop}%
\bibitem [{\citenamefont {Vidal}\ \emph
  {et~al.}(2009{\natexlab{a}})\citenamefont {Vidal}, \citenamefont {Dusuel},\
  and\ \citenamefont {Schmidt}}]{Vidal09_1}%
  \BibitemOpen
  \bibfield  {author} {\bibinfo {author} {\bibfnamefont {J.}~\bibnamefont
  {Vidal}}, \bibinfo {author} {\bibfnamefont {S.}~\bibnamefont {Dusuel}}, \
  and\ \bibinfo {author} {\bibfnamefont {K.~P.}\ \bibnamefont {Schmidt}},\
  }\href {\doibase 10.1103/PhysRevB.79.033109} {\bibfield  {journal} {\bibinfo
  {journal} {Phys. Rev. B}\ }\textbf {\bibinfo {volume} {79}},\ \bibinfo
  {pages} {033109} (\bibinfo {year} {2009}{\natexlab{a}})}\BibitemShut
  {NoStop}%
\bibitem [{\citenamefont {Vidal}\ \emph
  {et~al.}(2009{\natexlab{b}})\citenamefont {Vidal}, \citenamefont {Thomale},
  \citenamefont {Schmidt},\ and\ \citenamefont {Dusuel}}]{Vidal09_2}%
  \BibitemOpen
  \bibfield  {author} {\bibinfo {author} {\bibfnamefont {J.}~\bibnamefont
  {Vidal}}, \bibinfo {author} {\bibfnamefont {R.}~\bibnamefont {Thomale}},
  \bibinfo {author} {\bibfnamefont {K.~P.}\ \bibnamefont {Schmidt}}, \ and\
  \bibinfo {author} {\bibfnamefont {S.}~\bibnamefont {Dusuel}},\ }\href
  {\doibase 10.1103/PhysRevB.80.081104} {\bibfield  {journal} {\bibinfo
  {journal} {Phys. Rev. B}\ }\textbf {\bibinfo {volume} {80}},\ \bibinfo
  {pages} {081104(R)} (\bibinfo {year} {2009}{\natexlab{b}})}\BibitemShut
  {NoStop}%
\bibitem [{\citenamefont {Dusuel}\ \emph {et~al.}(2011)\citenamefont {Dusuel},
  \citenamefont {Kamfor}, \citenamefont {Or\'us}, \citenamefont {Schmidt},\
  and\ \citenamefont {Vidal}}]{Dusuel11}%
  \BibitemOpen
  \bibfield  {author} {\bibinfo {author} {\bibfnamefont {S.}~\bibnamefont
  {Dusuel}}, \bibinfo {author} {\bibfnamefont {M.}~\bibnamefont {Kamfor}},
  \bibinfo {author} {\bibfnamefont {R.}~\bibnamefont {Or\'us}}, \bibinfo
  {author} {\bibfnamefont {K.~P.}\ \bibnamefont {Schmidt}}, \ and\ \bibinfo
  {author} {\bibfnamefont {J.}~\bibnamefont {Vidal}},\ }\href {\doibase
  10.1103/PhysRevLett.106.107203} {\bibfield  {journal} {\bibinfo  {journal}
  {Phys. Rev. Lett.}\ }\textbf {\bibinfo {volume} {106}},\ \bibinfo {pages}
  {107203} (\bibinfo {year} {2011})}\BibitemShut {NoStop}%
\bibitem [{\citenamefont {Tupitsyn}\ \emph {et~al.}(2010)\citenamefont
  {Tupitsyn}, \citenamefont {Kitaev}, \citenamefont {Prokof'ev},\ and\
  \citenamefont {Stamp}}]{Tupitsyn10}%
  \BibitemOpen
  \bibfield  {author} {\bibinfo {author} {\bibfnamefont {I.~S.}\ \bibnamefont
  {Tupitsyn}}, \bibinfo {author} {\bibfnamefont {A.}~\bibnamefont {Kitaev}},
  \bibinfo {author} {\bibfnamefont {N.~V.}\ \bibnamefont {Prokof'ev}}, \ and\
  \bibinfo {author} {\bibfnamefont {P.~C.~E.}\ \bibnamefont {Stamp}},\ }\href
  {\doibase 10.1103/PhysRevB.82.085114} {\bibfield  {journal} {\bibinfo
  {journal} {Phys. Rev. B}\ }\textbf {\bibinfo {volume} {82}},\ \bibinfo
  {pages} {085114} (\bibinfo {year} {2010})}\BibitemShut {NoStop}%
\bibitem [{\citenamefont {Wu}\ \emph {et~al.}(2012)\citenamefont {Wu},
  \citenamefont {Deng},\ and\ \citenamefont {Prokof'ev}}]{Wu12}%
  \BibitemOpen
  \bibfield  {author} {\bibinfo {author} {\bibfnamefont {F.}~\bibnamefont
  {Wu}}, \bibinfo {author} {\bibfnamefont {Y.}~\bibnamefont {Deng}}, \ and\
  \bibinfo {author} {\bibfnamefont {N.}~\bibnamefont {Prokof'ev}},\ }\href
  {\doibase 10.1103/PhysRevB.85.195104} {\bibfield  {journal} {\bibinfo
  {journal} {Phys. Rev. B}\ }\textbf {\bibinfo {volume} {85}},\ \bibinfo
  {pages} {195104} (\bibinfo {year} {2012})}\BibitemShut {NoStop}%
\bibitem [{\citenamefont {Wegner}(1994)}]{Wegner94}%
  \BibitemOpen
  \bibfield  {author} {\bibinfo {author} {\bibfnamefont {F.}~\bibnamefont
  {Wegner}},\ }\href {\doibase 10.1002/andp.19945060203} {\bibfield  {journal}
  {\bibinfo  {journal} {Ann. Phys. (Leipzig)}\ }\textbf {\bibinfo {volume}
  {3}},\ \bibinfo {pages} {77} (\bibinfo {year} {1994})}\BibitemShut {NoStop}%
\bibitem [{\citenamefont {Knetter}\ and\ \citenamefont
  {Uhrig}(2000)}]{Knetter00}%
  \BibitemOpen
  \bibfield  {author} {\bibinfo {author} {\bibfnamefont {C.}~\bibnamefont
  {Knetter}}\ and\ \bibinfo {author} {\bibfnamefont {G.~S.}\ \bibnamefont
  {Uhrig}},\ }\href {\doibase 10.1007/s100510050026} {\bibfield  {journal}
  {\bibinfo  {journal} {Eur. Phys. J. B}\ }\textbf {\bibinfo {volume} {13}},\
  \bibinfo {pages} {209} (\bibinfo {year} {2000})}\BibitemShut {NoStop}%
\bibitem [{\citenamefont {Knetter}\ \emph {et~al.}(2001)\citenamefont
  {Knetter}, \citenamefont {Schmidt}, \citenamefont {Gr\"uninger},\ and\
  \citenamefont {Uhrig}}]{Knetter01}%
  \BibitemOpen
  \bibfield  {author} {\bibinfo {author} {\bibfnamefont {C.}~\bibnamefont
  {Knetter}}, \bibinfo {author} {\bibfnamefont {K.~P.}\ \bibnamefont
  {Schmidt}}, \bibinfo {author} {\bibfnamefont {M.}~\bibnamefont
  {Gr\"uninger}}, \ and\ \bibinfo {author} {\bibfnamefont {G.~S.}\ \bibnamefont
  {Uhrig}},\ }\href {\doibase 10.1103/PhysRevLett.87.167204} {\bibfield
  {journal} {\bibinfo  {journal} {Phys. Rev. Lett.}\ }\textbf {\bibinfo
  {volume} {87}},\ \bibinfo {pages} {167204} (\bibinfo {year}
  {2001})}\BibitemShut {NoStop}%
\bibitem [{\citenamefont {Kamfor}()}]{Kamfor13}%
  \BibitemOpen
  \bibfield  {author} {\bibinfo {author} {\bibfnamefont {M.}~\bibnamefont
  {Kamfor}},\ }\href@noop {} {}\bibinfo {note} {Ph. D. thesis, TU
  Dortmund/UPMC, available at:
  \href{http://hdl.handle.net/2003/30360}{http://hdl.handle.net/2003/30360}}\BibitemShut
  {NoStop}%
\bibitem [{\citenamefont {Xu}\ and\ \citenamefont {Moore}(2004)}]{Xu04}%
  \BibitemOpen
  \bibfield  {author} {\bibinfo {author} {\bibfnamefont {C.}~\bibnamefont
  {Xu}}\ and\ \bibinfo {author} {\bibfnamefont {J.~E.}\ \bibnamefont {Moore}},\
  }\href {\doibase 10.1103/PhysRevLett.93.047003} {\bibfield  {journal}
  {\bibinfo  {journal} {Phys. Rev. Lett.}\ }\textbf {\bibinfo {volume} {93}},\
  \bibinfo {pages} {047003} (\bibinfo {year} {2004})}\BibitemShut {NoStop}%
\bibitem [{\citenamefont {Xu}\ and\ \citenamefont {Moore}(2005)}]{Xu05}%
  \BibitemOpen
  \bibfield  {author} {\bibinfo {author} {\bibfnamefont {C.}~\bibnamefont
  {Xu}}\ and\ \bibinfo {author} {\bibfnamefont {J.~E.}\ \bibnamefont {Moore}},\
  }\href {\doibase 10.1016/j.nuclphysb.2005.04.003} {\bibfield  {journal}
  {\bibinfo  {journal} {Nucl. Phys. B}\ }\textbf {\bibinfo {volume} {716}},\
  \bibinfo {pages} {487} (\bibinfo {year} {2005})}\BibitemShut {NoStop}%
\bibitem [{\citenamefont {Kitaev}(2006)}]{Kitaev06}%
  \BibitemOpen
  \bibfield  {author} {\bibinfo {author} {\bibfnamefont {A.}~\bibnamefont
  {Kitaev}},\ }\href {\doibase 10.1016/j.aop.2005.10.005} {\bibfield  {journal}
  {\bibinfo  {journal} {Ann. Phys. (N.Y.)}\ }\textbf {\bibinfo {volume}
  {321}},\ \bibinfo {pages} {2} (\bibinfo {year} {2006})}\BibitemShut {NoStop}%
\bibitem [{\citenamefont {Jiang}\ \emph {et~al.}(2008)\citenamefont {Jiang},
  \citenamefont {Brennen}, \citenamefont {Gorshkov}, \citenamefont {Hammerer},
  \citenamefont {Hafezi}, \citenamefont {Demler}, \citenamefont {Lukin},\ and\
  \citenamefont {Zoller}}]{Jiang08}%
  \BibitemOpen
  \bibfield  {author} {\bibinfo {author} {\bibfnamefont {L.}~\bibnamefont
  {Jiang}}, \bibinfo {author} {\bibfnamefont {G.~K.}\ \bibnamefont {Brennen}},
  \bibinfo {author} {\bibfnamefont {A.~V.}\ \bibnamefont {Gorshkov}}, \bibinfo
  {author} {\bibfnamefont {K.}~\bibnamefont {Hammerer}}, \bibinfo {author}
  {\bibfnamefont {M.}~\bibnamefont {Hafezi}}, \bibinfo {author} {\bibfnamefont
  {E.}~\bibnamefont {Demler}}, \bibinfo {author} {\bibfnamefont {M.~D.}\
  \bibnamefont {Lukin}}, \ and\ \bibinfo {author} {\bibfnamefont
  {P.}~\bibnamefont {Zoller}},\ }\href {\doibase 10.1038/nphys943} {\bibfield
  {journal} {\bibinfo  {journal} {Nat. Phys.}\ }\textbf {\bibinfo {volume}
  {4}},\ \bibinfo {pages} {482} (\bibinfo {year} {2008})}\BibitemShut {NoStop}%
\end{thebibliography}
%merlin.mbs apsrev4-1.bst 2010-07-25 4.21a (PWD, AO, DPC) hacked
%Control: key (0)
%Control: author (8) initials jnrlst
%Control: editor formatted (1) identically to author
%Control: production of article title (-1) disabled
%Control: page (0) single
%Control: year (1) truncated
%Control: production of eprint (0) enabled
%

\end{document}